%% file: Main.tex
\documentclass[fleqn,usenatbib]{mnras}

\usepackage{newtxtext,newtxmath}

\usepackage[T1]{fontenc}

\DeclareRobustCommand{\VAN}[3]{#2}
\let\VANthebibliography\thebibliography
\def\thebibliography{\DeclareRobustCommand{\VAN}[3]{##3}\VANthebibliography}

\newcommand{\frb}{FRB 20230708A}
\newcommand{\hmshh}{\ensuremath{^\mathrm{h}}}
\newcommand{\hmsmm}{\ensuremath{^\mathrm{m}}}
\newcommand{\hmsss}{\ensuremath{^\mathrm{s}}}
\newcommand{\degree}{\ensuremath{^{\circ}}}
\newcommand{\arcm}{\ensuremath{^{\prime}}}
\newcommand{\arcs}{\ensuremath{^{\prime\prime}}}
\newcommand{\taninv}{\tan^{-1}}

\usepackage{graphicx}	
\usepackage{amsmath}	

\def\code#1{\texttt{#1}} 
\usepackage{subcaption}
\usepackage{makecell}
\usepackage{upgreek}
 
\title[A quasi-periodic FRB with unique morphology]{\frb, a quasi-periodic FRB with unique temporal-polarimetric morphology.}

\author[T. Dial et al.]{
T. Dial,$^{1}$\thanks{E-mail: tdial@swin.edu.au}
A. T. Deller,$^{1}$
P. A. Uttarkar, $^{1}$
M. E. Lower, $^{5}$
R.~M.~Shannon, $^{1}$
Kelly Gourdji, $^{1}$
\newauthor
Lachlan Marnoch, $^{3,4,5,6}$
A.~Bera, $^{2}$
Stuart D. Ryder, $^{3,4}$
Marcin Glowacki, $^{2}$
J. Xavier Prochaska $^{7,8,9}$
\\
$^{1}$Center for Astrophysics and Supercomputing, Swinburne University of Technology, P.O. Box 218, Hawthorn, Vic 3122 Australia
\\
$^{2}$International Centre for Radio Astronomy Research, Curtin University, Bentley, WA 6102, Australia 
\\
$^{3}$School of Mathematical and Physical Sciences, Macquarie University, NSW 2109, Australia
\\
$^{4}$Astrophysics and Space Technologies Research Centre, Macquarie University, Sydney, NSW 2109, Australia
\\
$^{5}$Australia Telescope National Facility, CSIRO Space \& Astronomy, Box 76 Epping, NSW 1710, Australia
\\
$^{6}$
The ARC Centre of Excellence for All-Sky Astrophysics in 3 Dimensions (ASTRO 3D)
\\
$^{7}$
Department of Astronomy and Astrophysics, University of California, Santa Cruz, CA 95064, USA
\\
$^{8}$
Kavli Institute for the Physics and Mathematics of the Universe, 5-1-5 Kashiwanoha, Kashiwa 277-8583, Japan
\\
$^{9}$
Division of Science, National Astronomical Observatory of Japan, 2-21-1 Osawa, Mitaka, Tokyo 181-8588, Japan
}

\date{Accepted XXX. Received YYY; in original form ZZZ}

\pubyear{2015}

\begin{document}
\label{firstpage}
\pagerange{\pageref{firstpage}--\pageref{lastpage}}
\maketitle

\begin{abstract}
\input{abstract}
\end{abstract}




\input{Introduction}

\input{Results}

\input{Discussion}

\input{Conclusion}

\input{other}



\bibliographystyle{mnras}
\bibliography{ref} 



\appendix

\input{appendix}

\bsp	
\label{lastpage}
\end{document}

%% file: abstract.tex
There has been a rapid increase in the known fast radio burst (FRB) population, yet the progenitor(s) of these events have remained an enigma. A small number of FRBs have displayed some level of quasi-periodicity in their burst profile, which can be used to constrain their plausible progenitors.  
However, these studies suffer from the lack of polarisation data which can greatly assist in constraining possible FRB progenitors and environments. Here we report on the detection and characterisation of \frb\ by the Australian Square Kilometre Array Pathfinder (ASKAP), a burst which displays a rich temporal and polarimetric morphology. We model the burst time series to test for the presence of periodicity, scattering and scintillation. 
We find a potential period of T = 7.267 ms within the burst, but with a low statistical significance of 1.77$\sigma$.
Additionally, we model the burst's time- and frequency-dependent polarisation to search for the presence of (relativistic and non-relativistic) propagation effects.
We find no evidence to suggest that the high circular polarisation seen in \frb\ is generated by Faraday conversion.
The majority of the properties of \frb\ are broadly consistent with a (non-millisecond) magnetar model in which the quasi-periodic morphology results from microstructure in the beamed emission, but other explanations are not excluded.

%% file: Introduction.tex
\section{Introduction}

Fast radio bursts (FRBs) are bright (sub-)millisecond bursts of radio emission which possess many similarities with radio emission from Galactic neutron stars (NSs). However, their extragalactic origins require luminosities that are $\geq$9-10 orders of magnitude larger than what is typically observed for Galactic NSs \citep{petroff2019fast, petroff2022fast} and as such their progenitors have remained an enigma.
Recent studies \citep{gordon2023demographics,law2024deep,shannon2024commensal} have looked at the host galaxies of FRBs to constrain the potential pathways of FRB progenitors. However, of the O(1000) published FRBs, only $\sim 50$ have corresponding host galaxies due to the necessity of (sub-)arcsecond localisation and follow up through deep optical imaging. From the limited sample of FRB host galaxy observations, it has been shown that the larger population tends to track the star-forming main sequence of galaxies \citep{bhandari2022characterizing, gordon2023demographics}, which could favour progenitor models with short delay channels such as magnetars formed via core-collapse supernovae. However, some individual FRBs contradict this picture, most notably
FRB 20200120E, a repeating FRB localised to a globular cluster of the spiral galaxy M81 \citep{kirsten2022repeating}.
The potential for multiple progenitors is further implied by the seemingly dissimilar properties between the two observed populations of FRBs, those that are proven to repeat, and one-off FRBs yet to be observed to repeat. Repeating FRBs in general show broader temporal profiles and narrower bandwidths in their emission compared to their non-repeating counterparts \citep{amiri2021first, pleunis2021fast}. Some repeaters have also been shown to exist in extreme magneto-ionic environments, such is the case for FRB 20121102A with its high apparent rotation measure (RM) variability \citep{michilli2018extreme, hilmarsson2021rotation, plavin2022frb}, and FRB 20190520B which has been observed to undergo a magnetic field reversal \citep{anna2023magnetic}. Progenitor models involving a NS embedded in a turbulent environment such as a supernova remnant or in the proximity of a black hole \citep{michilli2018extreme, hilmarsson2021rotation} and NS binary systems \citep{anna2023magnetic} have been suggested to explain these observations.

The most direct evidence linking FRBs to a potential progenitor is the bright radio burst detected from the Galactic magnetar SGR J1935+2154 in 2020 \citep{andersen2020bright, Prog:bochenek2020fast}. The brightest burst showed striking similarities to the known FRB population. The >1\,MJy\,ms fluence was orders of magnitude more energetic than previous magnetar radio emission, but still at least 2 orders of magnitude less energetic than the faintest non-repeating FRB \citep{law2024deep} \cite[although of comparable brightness to the least energetic bursts observed from nearby repeating FRBs with deep follow up;][]{nimmo2023burst}. The burst also showed FRB-like morphology with millisecond-scale sub-components. Analysis of this burst, and subsequent bursts detected after it have also suggested FRBs to be the byproducts of magnetosphere instabilities \citep{zhang2020highly,kirsten2021detection,zhu2023radio}. However, despite the belief that at least a sub-set of FRBs may originate from magnetars, the possibility of multiple progenitor pathways is still open. For instance, the apparent one-off nature of some FRBs may be evidence of FRBs originating from cataclysmic progenitors such as the merger of compact objects \cite[e.g.][]{totani2013cosmological}.

FRB progenitor models can be tested by taking advantage of the diversity of temporal, spectral and polarimetric properties they exhibit. The presence of multiple components within a single burst \citep{amiri2021first, sherman2024deep}, 
high linear polarisation (LP) fractions \citep{day2020high} \citep[and occasionally circular polarisation (CP);][]{pandhi2024polarization}, and the behaviour of the observed linear polarisation position angle (PA) (often flat, but sometimes smoothly or erratically varying within a pulse) all constrain potential progenitors. All these properties have been observed in Galactic magnetars, but none are exclusive to magnetars.

Propagation effects observed in FRBs can also constrain possible progenitors. Scattering and scintillation studies of FRBs using the double thin-screen model have been used to constrain the locations of dominant turbulent media along FRB sight lines, providing insight into the circum-galactic medium or environments local to the FRB source \citep{ Scin:masui2015dense, ocker2022large, sammons2023two}. Such local environments may be directly involved in producing the high CP seen in some FRBs, possibly due to propagation through a highly relativistic plasma \citep{kumar2023propagation} or through a maser mechanism model \citep{faber2023morphologies}.

A handful of FRBs have also shown quasi periodicity within their burst structure \citep{chime2022sub}, the most significant detection of which, with a period of T$\simeq$216.8~ms, could be explained by beamed emission from a rotating pulsar or magnetar \citep{chime2022sub}. Other possible models such as direct-current circuit breaking in compact mergers \citep{chime2022sub} and crustal oscillations on a magnetar surface \citep{Pastor_Marazuela_2023, wadiasingh2020fast} have also been theorised. \citet{kramer2023quasi} also showed that the quasi-periodic sub-structure found in the beamed emission of pulsars and magnetars and their rotational period follow a direct scaling relationship. If some FRBs do come from NS's, the sub-structure in FRBs could be used to infer their underlying periods. However, no direct evidence has been reported to date.

Here we report on \frb, an apparently non-repeating FRB that exhibits a plethora of striking temporal, spectral and polarimetric properties. In Section \ref{sec:METHODS} we briefly discuss the detection of \frb\ and present an analysis of its properties, including morphology, scattering, scintillation and polarisation. In Section \ref{sec:DISCUSSION} we comment on the potential quasi-periodicity and unusually high CP, and discuss the possible progenitors and sources that could produce \frb. Finally, we present our conclusions in Section \ref{sec:CONCLUSION}. Modelling performed in this study (unless specified otherwise) made use of Bayesian inference with the Dynesty (Dynamic nested sampling) Sampler \citep{speagle2020dynesty} using \textsc{Bilby} \citep{ashton2019bilby}. Unless otherwise stated, all uncertainties are reported to 68$\%$ confidence.

%% file: Results.tex
\section{Method \& Results}
\label{sec:METHODS}

\subsection{Detection/localisation of \frb}

On 2023 Jul 08 at UTC 15:32:47, the incoherent summation (ICS) FRB detection system on ASKAP \citep{hotan2021australian} reported the discovery of an FRB \citep{shannon2024commensal} as part of the Commensal Real-Time ASKAP Fast-Transients collaboration \citep[CRAFT;][]{macquart2010commensal}. The FRB was detected with a dispersion measure (DM) of 411.12 pc\,cm$^{-3}$ and width of 4 ms whilst ASKAP was observing in the ``low" band: with  burst search system observations having a  central frequency of 919.5 MHz with 336 MHz of bandwidth. The detection had a latency of 0.94s from the voltage download trigger at a frequency of 751.5 MHz, which combined with a dispersive delay sweep of 1.58\,s meant that the FRB emission was captured across the same  336 MHz band within the 3.1\,s of raw voltage data. A download of the voltage data was triggered by the real-time detection system and transferred to the Ngarrgu Tinderbeek supercomputer at Swinburne University of Technology. After the active ASKAP observations had concluded, calibration observations of the polarisation calibrator PSR J0835$-$4510 (Vela) and bandpass calibrator, the radio galaxy PKS B0407$-$658, were taken 8.15 and 8.30 hours respectively after the initial FRB detection, with 3.1\,s of raw voltage data similarly obtained for each source. The raw voltages for the FRB and calibrator sources were then processed using the CRAFT Effortless Localisation and Enhanced Burst Inspection (CELEBI) pipeline \citep{scott2023celebi} to produce an accurate sub-arcsecond position and structure maximised DM (reported in Table~\ref{tab:frbparam}) for \frb. This was used to produce coherently de-dispersed beamformed high time resolution (HTR) dynamic spectra with full polarimetry.

The CELEBI pipeline makes use of a polarisation calibrator observation to correct for instrumental polarisation leakage, as well as any rotational offset between the antenna and sky coordinate system. We have implemented improvements to the polarisation calibration code previously reported in \citet{scott2023celebi} by solving for the three polarisation calibration terms described in Eq 9-11 of the supplementary material of \citet{bannister2019single} using the Dynesty sampler \citep{speagle2020dynesty} enabled by Bilby \citep{ashton2019bilby}. Briefly, we model Vela with a constant linear and circular polarisation fraction of $l$ = L/I = 0.95 and $v$ = V/I = --0.05 respectively across frequency \citep{johnston2018polarimetry}. We also model the linear polarization position angle (PA) to be 0.35 rad at 1400 MHz \citep{bannister2019single}. We sample the polarisation leakage parameters $\tau$ and $\phi$ describing the time and phase delay between the X and Y polarisation voltages, the rotation offset of the antenna phased array feed (PAF) $\psi$, the total integrated rotation measure (RM) as well as linear and circular polarisation scale factors L$_{\rm scale}$ and V$_{\rm scale}$ to account for zeroth order variations in polarisation over different observing bands. The parameter priors and posteriors for Vela are reported in Table~\ref{tab:velaparams}.

\begin{table}
    \centering
    \caption{Priors and posteriors for Vela parameters in the polarisation calibration workflow.}
    \begin{tabular}{l|c|c}
        \hline
        Parameter & Prior & Posterior \\
        \hline
        $\tau$ (ns)                  & [--1000, 1000]       & 99 $\pm$ 6  \\
        $\phi$ (rad)                 & [$-\pi$, $\pi$]       & 0.57 $\pm$ 0.04   \\
        $\psi$ (rad)                 & [$-\pi$/2, $\pi$/2]   & 0.748 $\pm$ 0.004  \\
        RM (rad m$^{-2}$)            & [30, 50]              & 38.79 $\pm$ 0.06  \\
        L$_{\mathrm{scale}}$         & [0.7, 1.2]            & 0.992 $\pm$ 0.003  \\
        V$_{\mathrm{scale}}$         & [--1.0, 1.2]           & 0.95 $\pm$ 0.05  \\
        \hline
    \end{tabular}
    
    \label{tab:velaparams}
\end{table}

Imaging observations of the field surrounding the position of \frb\ were made using the European Southern Observatory's Very Large Telescope (VLT; Project ID 108.21ZF, PI Shannon) on 2023~July~21 with the FORS2 instrument for 2000~s in the $R$-band; and on 2023 July 26 with the HAWK-I instrument + GRAAL ground-layer adaptive optics module for 2400~s in the $K$-band. The images were processed in the manner described by \citet{Marnoch2023}.
The well-constrained position of \frb\ is coincident with a galaxy and far from any other sources, as shown in Fig.~\ref{fig:hawk}. Using the PATH package \citep{aggarwal2021probabilistic} and VLT $R$-band image (shown in the left panel of Fig.~\ref{fig:hawk}), we securely identified this galaxy to be the host (with a false association probability of $8 \times 10^{-7}$). We proceeded with acquisition of a spectroscopic redshift for this galaxy. Observations with the X-shooter instrument on the VLT on 2023~Aug~7 were made in sub-arcsecond seeing, showing clear detections of [O\,{\sc ii}] $\lambda\lambda$3726,3729, [O\,{\sc iii}] $\lambda\lambda$4959,5007, H$\beta$ and H$\alpha$ emission lines consistent with a redshift of $z = 0.1050 \pm 0.0001$ \citep[][Muller et al. in prep.]{shannon2024commensal}.

\begin{table}
    \centering
    \caption{Properties of \frb. $\tau_{s}$ is the scattering timescale (at the central frequency of 919.5 MHz), $\nu_{dc}$ is the decorrelation bandwidth, $\nu_{\mathrm{NE2001}}$ is the decorrelation bandwidth based on the NE2001 model \citep{cordes2002ne2001}, $C$ is the scintillation constant (Eq.\ref{eq:scint_constant}), $\alpha_{t}$ is the scattering index, $z$ is the host galaxy redshift, $\rm L_{g}$ is the distance between earth and major galactic thin scattering screen, $\rm L_{x}$ is the distance between the FRB source and host galaxy major thin scattering screen, $\rm m$ is the modulation index and $\rm T$ is the (quasi-)periodicity in the burst. $\rm{MJD}$ reported here is the peak of the Stokes $I$ time series of the burst at a reference frequency of 751.5 MHz using the geocentric reference frame of the earth. The uncertainty in the fluence is estimated to be 10\%, dominated by our ability to model gain variations in the ASKAP bandpass across frequency regions afflicted by RFI on the calibrator scan. The total energy and peak luminosity were calculated assuming the $\Lambda$-CDM model with cosmological parameters derived from \citet{aghanim2020planck}. The integrated linear, absolute circular and total polarization fractions ($\overline{l}$, $\overline{|v|}$ and $\overline{p}$) were calculated using Eq.\ref{eq:int_polfrac} by integrating over the first bright sub-burst (Fig.~\ref{fig:comp1_fit}). The peak linear and circular polarisation fractions, $l_{\rm peak}(t)$ and $|v|_{\rm peak}(t)$ were estimated at a time resolution of 100 $\mu s$ and masking anything above a S/N = 3. $l_{\rm peak}(t)$ and $|v|_{\rm peak}(t)$ are found at a time offset of -0.71 ms and 0.59 ms from the peak of the burst respectively.}
    \begin{tabular}{ l | c }
    \hline
        Parameters                      & Derived values                \\         
    \hline
        RA (J2000)                            & 20\hmshh12\hmsmm27\hmsss.73 $\pm$ 0.47  \\
        DEC (J2000)                            & -55\degree21\arcm22\arcs.6 $\pm$ 0.44 \\
        DM  (pc cm$^{-3}$)              & 411.51 $\pm$ 0.05                        \\
        $\rm{MJD}$ (days)                      & 60133.647766200826              \\
        $\tau_{s}$  (ms)                & 0.17 $\pm$ 0.02               \\
        $\nu_{dc}$  (MHz)               & 0.38 $\pm$ 0.07               \\
        $\nu_{\mathrm{NE2001}}$  (MHz)           & 0.43                        \\
        C                               & 395 $\pm$ 70                          \\
        $\alpha_{t}$                        & -2.17$_{-0.08}^{+0.12}$       \\
        z                               & 0.1050 $\pm$ 0.0001                         \\
        L$_{g}$L$_{x}$  (kpc$^{2}$)     & $\lesssim$ 97 $\pm$ 18                \\
        L$_{x}$ (kpc)                   & $\lesssim$ 62 $\pm$ 13                    \\
        m                               & 0.31 $\pm$ 0.07               \\
        RM  (rad m$^{-2}$)              & -6.90 $\pm$ 0.04              \\
        T  (ms)                         & 7.267                         \\
        burst width (ms)                & 26.44                         \\
        Fluence (Jy ms)                 & 89.2 $\pm$ 0.9                \\
        Total Energy (ergs)             & (9.1 $\pm$ 1.0)$\times$10$^{39}$ \\
        Peak Luminosity (ergs s$^{-1}$)        & (8.1 $\pm$ 0.9)$\times$10$^{40}$ \\
        $\overline{l}$                               & 0.684 $\pm$ 0.006             \\
        $\overline{|v|}$                             & 0.401 $\pm$ 0.005             \\
        $\overline{p}$                               & 0.878 $\pm$ 0.006             \\
        $l_{\rm peak}$                  & 0.99 $\pm$ 0.02               \\
        $v_{\rm peak}$                  & 0.79 $\pm$ 0.09               \\
        
        \hline
        
    \end{tabular}
    
    \label{tab:frbparam}
\end{table}

\subsection{Total intensity time series analysis}
\label{sec:time}

\begin{figure*}
    \centering
    \includegraphics[width = 0.9\textwidth]{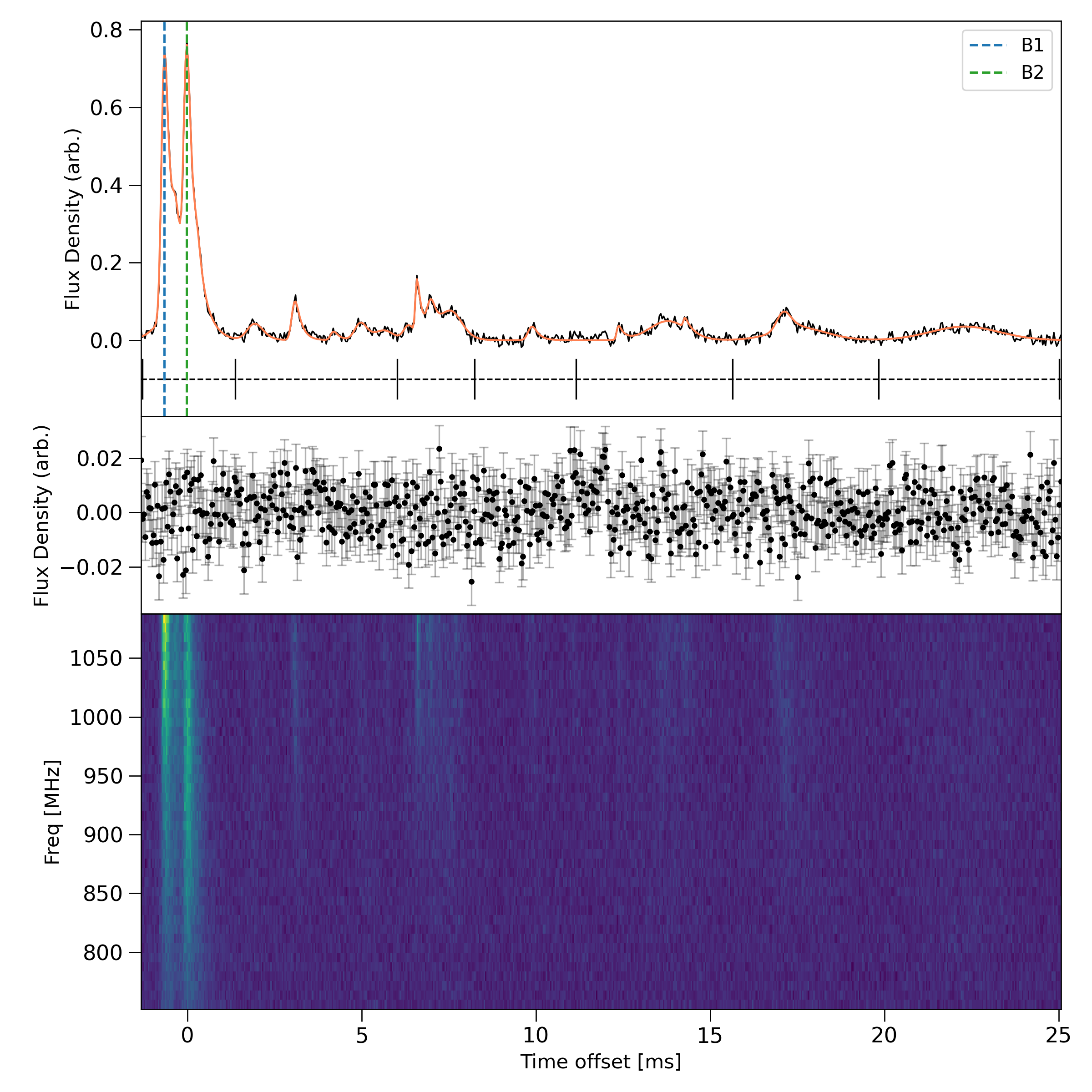}
    \caption{ \\FRB\ profile and dynamic spectrum.  Top Panel: Stokes $I$ burst profile with full burst model shown in orange which consists of 21 Gaussian pulses convolved with a one-sided exponential. The black dashed lines underneath the burst profile illustrate the 7 segments the burst was separated into for fitting the full burst. As discussed in Section~\ref{sec:GFR}, the two largest peaks in the first bright sub-burst are labelled B1 and B2. Middle Panel: Residuals of burst fitting. Bottom Panel: Stokes $I$ dynamic spectrum. The  burst data has been further averaged to $\rm 40 \mu s$ time resolution 8 MHz frequency resolution for visual aid. The time offset on the X-axis is relative to the burst MJD reported in Table~\ref{tab:frbparam}.}
    \label{fig:burstfit}
\end{figure*}

\frb\ has a complicated and striking temporal morphology. The stokes $I$ dynamic spectrum and time series burst profile in Fig.~\ref{fig:burstfit} show a multi-component burst with quasi-periodic sub-bursts successively decreasing in brightness over the full burst duration. We define any emission which is distinct and separated from other emission as a "sub-burst".
The data used to fit the burst model was the Stokes $I$ dynamic spectrum which was frequency averaged across the full 336 MHz bandwidth and down-sampled to a time resolution of 10$\mu$s. We model each sub-burst using one or more "components", where a component is a single Gaussian (whose width, amplitude and central time are free to vary) convolved with a one-sided exponential tail (whose properties are held constant across all components) with a timescale $\tau_{s}$ (adapted from Eq.4 of \citet{qiu2020population})

\begin{equation}
    I(t) = \sum_{i = 1}^{N}\bigg[A_{i}e^{-(t-\mu_{i})^{2}/2\sigma_{i}^{2}}\bigg] * e^{-t/\tau_{s}}.
    \label{eq:gauss}
\end{equation}

\begin{figure}
    \centering
    \includegraphics[width = \linewidth]{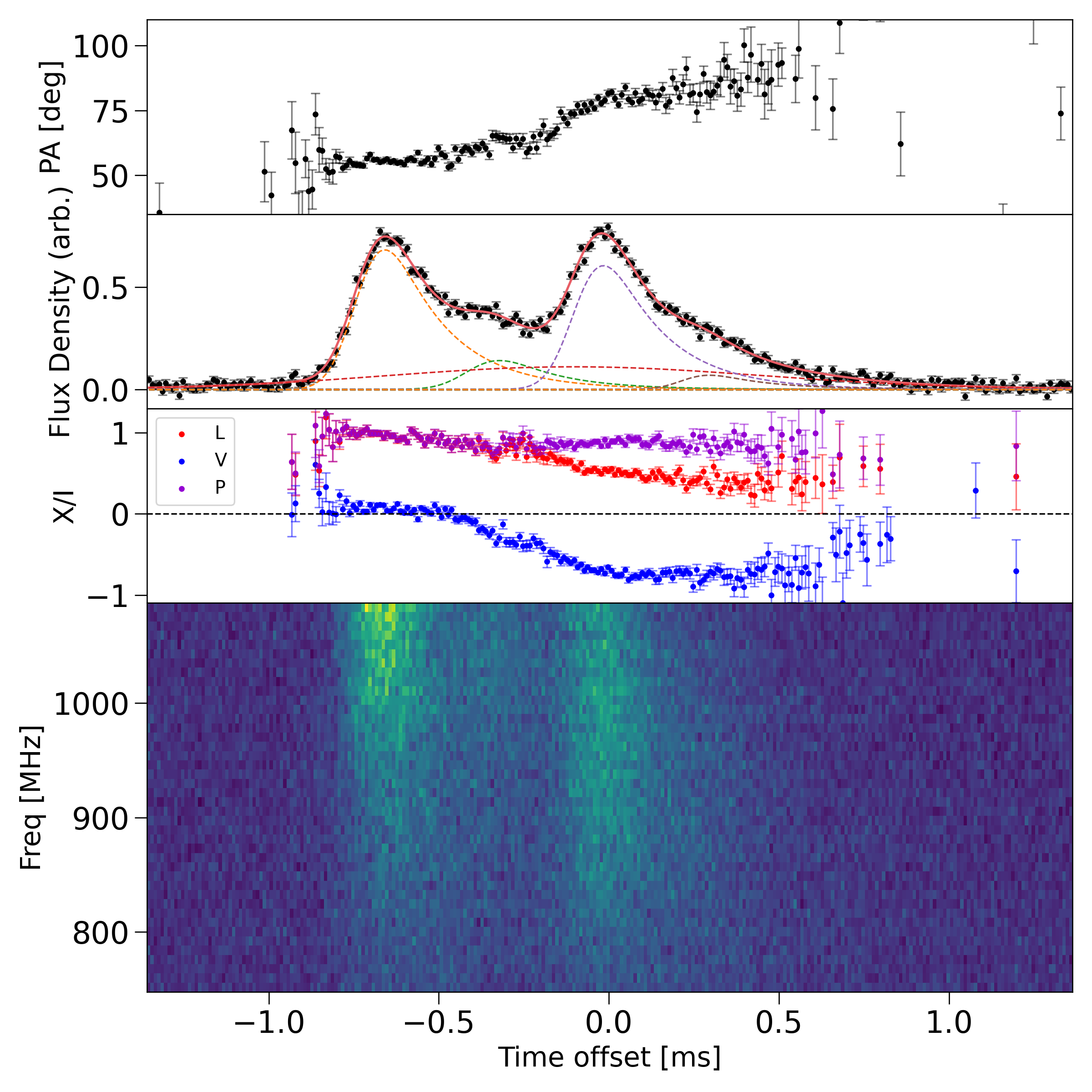}
    \caption{Polarimetry of the first sub-burst.  Panel 1: PA profile. Panel 2: Stokes $I$ time-series profile showing model fit as a bold orange line. Each Gaussian pulse is convolved with the same exponential and shown as a dashed line of a different color. Panel 3: $L$/$I$ in red, $V$/$I$ in blue and total polarisation $P$/$I$ in purple. Panel 4: Stokes $I$ dynamic spectrum with 8 MHz frequency resolution for visual aid.}
    \label{fig:comp1_fit}
\end{figure}

Where A$_{i}$, $\mu_{i}$, and $\sigma_{i}$ are, respectively, the amplitude, position in time and pulse width (standard deviation of the Gaussian pulse) in time of each $i^{\mathrm{th}}$ Gaussian, and $'*'$ denotes the convolution operation. 
We chose to divide the profile into seven distinct segments to simplify and speed up modelling (see Fig.~\ref{fig:burstfit}). Assuming negligible temporal variations in the scattering time scale $\tau_{s}$ across few millisecond duration of the burst, we modelled the first segment to constrain $\tau_{s}$ given the much higher S/N. We applied an iterative process of adding Gaussian pulses together to constrain a model of this segment. For each pulse, starting from one pulse, we sampled the position across the extent of the segment, the width up to the half width half maximum (HWHM) of the segment, the amplitude up to the peak of the segment, and $\tau_{s}$ up to 1\,ms. Once constrained, each new iteration of the model added an additional pulse with these wide priors. This process terminated when the Bayesian Information Criterion  \citep[BIC;][]{neath2012bayesian} ceased increasing. The first segment of \frb\ in Fig.~\ref{fig:comp1_fit} was found to be well described by the sum of five individual Gaussian pulses with a scattering timescale of $\tau_{s}$ = 0.17 $\pm$ 0.02 ms. We fitted the other segments independently and found $\tau$ to be consistent across the burst (but with much lower precision); accordingly, we subsequently modelled the remaining 6 segments with $\tau_{s}$ fixed to this value to minimise covariance with the other fitted parameters. The full burst model is shown in Fig.~\ref{fig:burstfit}.

We also investigated the frequency dependence of the scattering. 
The scattering index $\alpha_{t}$ describing how the scattering timescale scales with frequency assuming a power-law function $\tau_{s} \propto f^{\alpha_{t}}$, where f is in MHz, was also modelled following a similar procedure in \citet{qiu2020population,sammons2023two}, whereby the first sub-burst in the Stokes $I$ dynamic spectrum was split into five 67.2 MHz sub-bands. Each sub-band was down-sampled to 10$\mu$s resolution and modelled assuming a sum of five convolved Gaussian pulses with fixed positions and widths derived from the full band model described above and in Fig.~\ref{fig:comp1_fit}; modelling was also done with just the burst centroids fixed, allowing the pulse widths to vary. However, no significant differences in the width were found. Fig.~\ref{fig:specindex} shows the power law function fitted to the five width measurements (and their uncertainties) which results in a scattering index of $\alpha_{t}$ = -2.17$_{-0.08}^{+0.12}$.

\subsection{Periodicity}

\begin{figure}
    \centering
    \includegraphics[width = \linewidth]{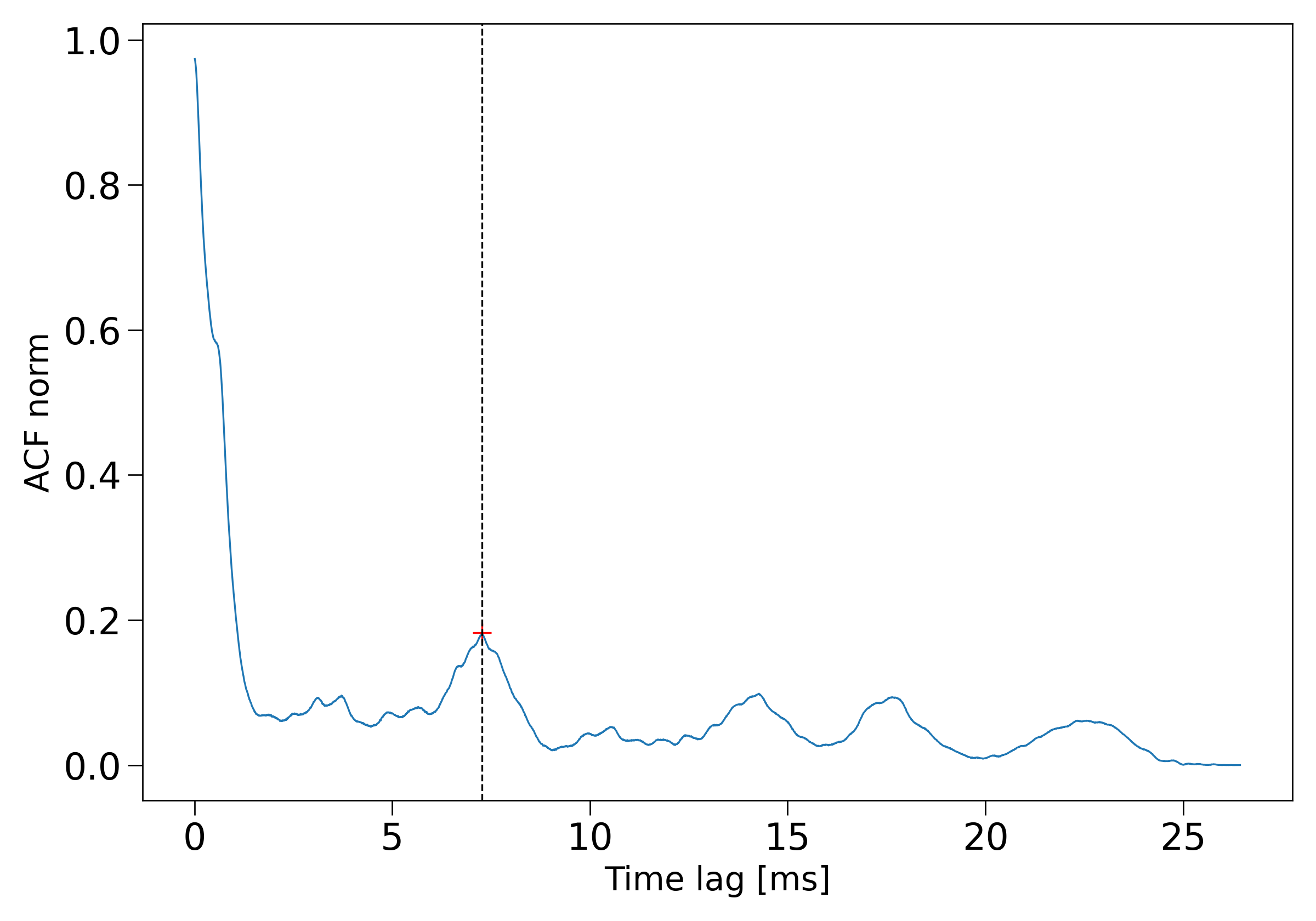}
    \caption{Normalised ACF of full burst. The red dot and black dashed line indicate the best fit for periodicity.}
    \label{fig:time_acf}
\end{figure}

Motivated by the visual suggestion of periodicity in the time series of the full burst, we undertook periodicity searches following the approaches used by \citet{chime2022sub} and \citet{Pastor_Marazuela_2023}. The auto-correlation function (ACF) of the full burst is shown in Fig.~\ref{fig:time_acf}. Assuming the first brightest peak in the ACF away from the non-zero time lag is the best fit for a periodicity, we estimated the burst period to be T = 7.267\,ms. The bottom panel of Fig.~\ref{fig:PA_plot} illustrates this potential periodicity with vertical dashed lines on the time series.

Next, we attempted to measure the statistical significance of T. One important aspect of \frb\ is that some of the sub-bursts have detailed sub-structure on $\mu$s timescales. Previous FRBs exhibiting (quasi-)periodicity \citep{chime2022sub, Pastor_Marazuela_2023} in contrast, could be modelled adequately using a single Gaussian to represent each pulse component. It may be the case that either the time resolution or S/N has been insufficient to discern such features, if they were present, and as such the treatment of resolved sub-structure in (quasi-)periodic emission is not something that has previously been necessary. Our approach to the presence of sub-structure was to extract a single mean time, effective width and amplitude for each sub-burst by modelling it as if it were a single component. This new burst envelope, now made up of 11 sub-bursts, was used to test the statistical significance of T. 

\begin{figure}
    \centering
    \includegraphics[width = \linewidth]{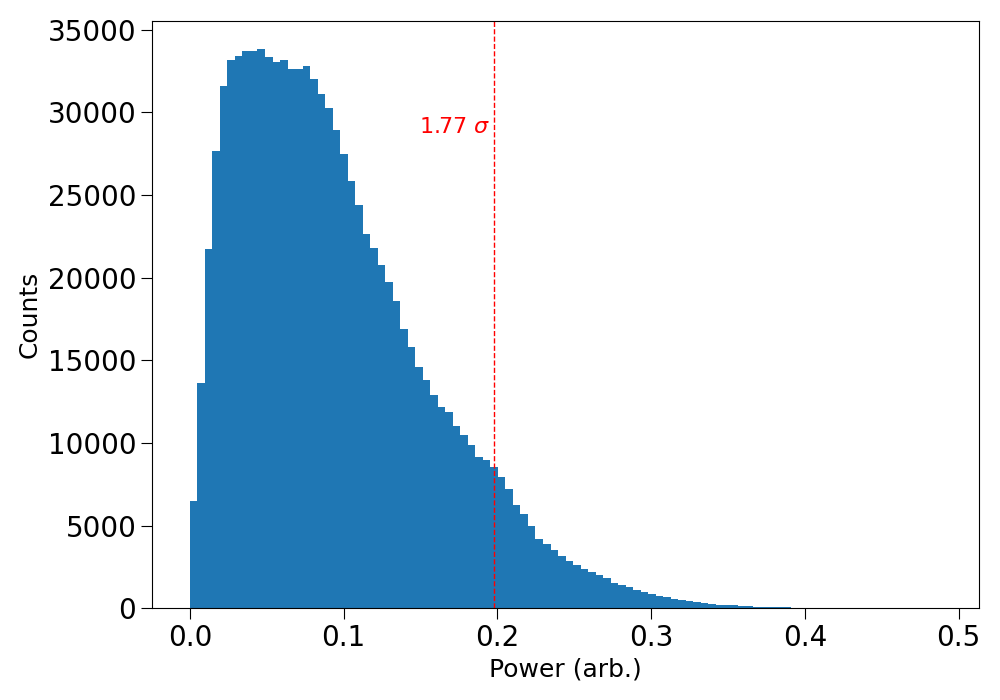}
    \caption{Figure shows the distribution of null hypothesis tests (in blue) according to the ACF power test \citet{kramer2023quasi}. The red dashed line shows the ACF power score of the original burst envelope with a statistical significance of 1.77 $\sigma$.}
    \label{fig:sigtest}
\end{figure}

We made use of the ACF power test described in \citet{kramer2023quasi}. This test was chosen due to the imprecise periodicity shown in the burst, for which other tests such as the Rayleigh test would be less optimal (for additional details, see the supplementary material of \citealt{kramer2023quasi}). The test involves measuring the ACF of the burst envelope from a time lag of zero to half the burst width. The points along the ACF at intervals of T were added together to produce an ACF power score. To estimate the statistical significance we simulated $10^{6}$ null hypothesis tests. For each test, the positions of each pulse in the burst envelope were scrambled with an average separation $\bar{d}$ = T and random variation drawn from a probability distribution (see Eq.~9 of \citealt{chime2022sub}, where the dimensionless parameter $\chi$ = 0.2) to produce a new burst envelope. The ACF power score was calculated for each test, the distribution of which, shown in Fig.~\ref{fig:sigtest}, was used to estimate a statistical significance of 1.77 $\sigma$ for periodicity in \frb. While visually rather striking, the apparent periodicity in \frb\ is thus only suggestive, rather than definitive.

\subsection{Polarisation Analysis}
\label{sec:pol}

The complex temporal structure of \frb\ complicates any polarimetric analysis. We began by estimating the mean RM across the burst. Some one-off FRBs, such as FRB 20221101B and FRB 20220207C, have shown significant apparent RM variations of up to $3\sigma$ over $\leq$5ms \citep{sherman2024deep} across their pulse profiles. Such apparent RM variation is also observed in radio pulsars \cite[][]{dai2015msps, ilie2019psrs}. To account for any possible RM variation over the $\sim30$~ms burst of \frb\ we split the burst into uniform 3.3 ms chunks and scrunched in time to form spectra from the 1MHz resolution Stokes $Q$ and $U$ dynamic spectra, using the burst model as a matched filter when averaging. To fit the RM of each chunk we used RM synthesis using the \code{RMtools} python package \citep{2020ascl.soft05003P}.

 Fig.~\ref{fig:RM_burst} shows the fitted RM across the burst profile. The uncertainties have been scaled by$\sqrt{\chi^2_{r}}$, where $\chi^2_{r}$ is the reduced chi-square of the fit. The RM of the burst is $\sim -7$~rad/m$^{2}$ across the first bright sub-burst, where it can be precisely measured. Moderate variations around this value (up to $-$13~rad/m$^{2}$) can be seen in the fainter sub-bursts, however, we note that these variations are of marginal significance ($\sim2\sigma$). Given that the formal RM uncertainties are likely underestimated given the scaling noted above, we consider this evidence of RM variation to be tentative at best. Hence, we assume a constant RM across the full burst, which was measured by applying the full burst fit shown in Fig.~\ref{fig:burstfit} as a matched filter in time to form Stokes $I$, $Q$ and $U$ spectra with 1MHz frequency resolution. Using RM synthesis we derived an RM of $-6.90 \pm 0.04$~rad/m$^{2}$. In Fig.~\ref{fig:RM_full} we plot the measured frequency dependent PA:
  \begin{equation}
    \mathrm{PA_{data}(\nu)} = \frac{1}{2}\mathrm{tan}^{-1}\bigg(\frac{U(\nu)}{Q(\nu)}\bigg).   
    \label{eq:PA}
\end{equation} 

We then compare the measured PA, $\rm PA_{data}$, to the expected model

\begin{equation}
    \mathrm{PA_{model}(\nu) = RMc^{2}}\bigg(\frac{1}{\nu^{2}} - \frac{1}{\nu_{0}^{2}}\bigg),
    \label{eq:RM}
\end{equation}

\noindent
to test if the data is well described by the modelled RM. 
In the above model, $\nu_{0}$ is the reference frequency, which is calculated as the weighted average of the wavelength squared values of the channels \citep{brentjens2005faraday}. The RM was then used to remove the Faraday rotation from the Stokes $Q$ and $U$ dynamic spectrum following

 \begin{equation}
    \begin{split}
        & Q\mathrm{_{deRM}} = Q\mathrm{cos(2PA_{model})} + U\mathrm{sin(2PA_{model})} \\
        & U\mathrm{_{deRM}} = Q\mathrm{sin(2PA_{model})} - U\mathrm{cos(2PA_{model})}. \\
    \end{split}
    \label{eq:FdayRot}
\end{equation}

\begin{figure*}
    \centering
    \includegraphics[width = 0.9\textwidth]{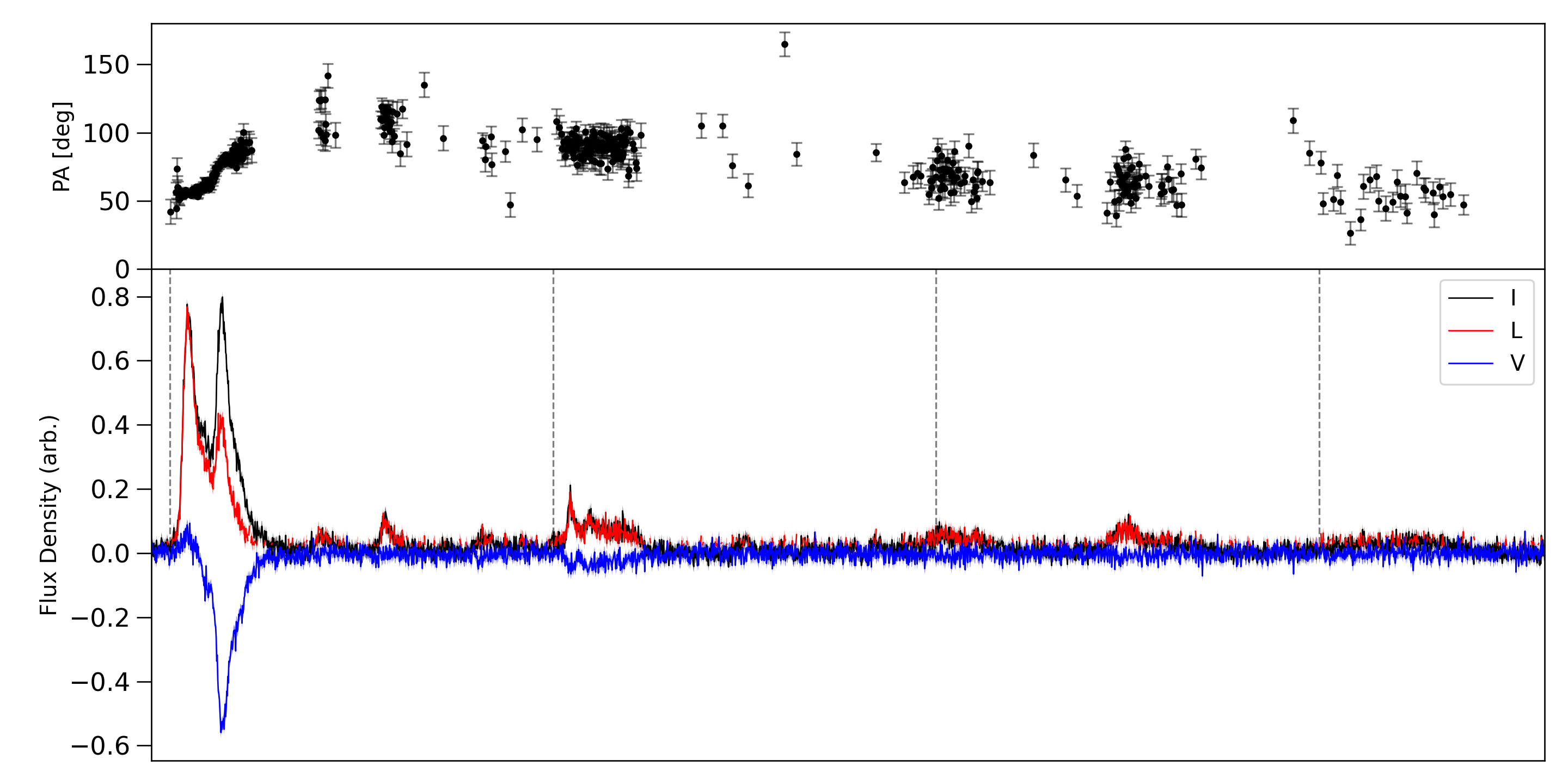}
    \caption{Top: PA profile of full burst with 10 $\mu$s time resolution. Bottom: Stokes time series. A periodicity of T = 7.267ms is plotted against the timeseries using evenly spaced grey dashed vertical lines starting from the first sub-burst. }
    \label{fig:PA_plot}
\end{figure*}

We then calculated the de-biased linear polarisation fraction L$_\mathrm{debias}$ \citep{everett2001emission,day2020high}, as shown on the bottom panel of Fig.~\ref{fig:PA_plot}, using
 
 \begin{equation}
    L\mathrm{_{debias}} = 
    \begin{cases} 
      \sigma_{I}\sqrt{\bigg(\frac{L}{\sigma_{I}}\bigg)^{2} - 1} & \frac{L}{\sigma_{I}} > 1.57 \\
      0 & \mathrm{otherwise}. \\
   \end{cases}
   \label{eq:L_debias}
\end{equation}
 
 Eq.~\ref{eq:PA} was used to calculate the time-dependent PA across the full burst, which is only reported when the linear polarization fraction exceeds the threshold
\begin{equation}
    \mathrm{PA(t)} = 
    \begin{cases}
        \mathrm{PA(t)} & L\mathrm{_{debias}} \geq \mathrm{B}\sigma_{I} \\
         $--$ & \mathrm{otherwise}. \\
    \end{cases}
    \label{eq:PA_debias}
\end{equation}
 
 The free parameter $\rm B$ in Eq.~\ref{eq:PA_debias} is a tuneable parameter used to control the error threshold for plotting the PA. For this study we chose $\rm B$ = 3.0. 
 The resulting PA plot is shown in the top panel of Fig.~\ref{fig:PA_plot}.

We attempted to fit a rotating vector model \citep[RVM][]{rvm}  , with a period of T = 7.267 ms, to the full burst PA profile to explore the potential to a rotation powered NS such as a pulsar. However, the results were non-constraining, providing no evidence for or against a RVM.

\subsection{Searching for Generalised Faraday Rotation in \frb}
\label{sec:GFR}
As shown in Fig.~\ref{fig:comp1_fit}, the first bright sub-burst of \frb\ shows extremely high CP ($V$/$I \sim 75\%$) in the trailing component. To date it is unclear what causes CP in FRBs. It is possible that small fractions of CP are intrinsic to the FRB emission \citep{zhang2023fast}, which may hold for the large fraction of FRBs that show low to moderate levels of CP ($V$/$I \leq 30\%$) \citep{pandhi2024polarization, sherman2024deep}. However, intrinsic emission alone can not explain the much higher CP fractions reported in a handful of FRBs \citep{feng2022circular,zhang2023fast,anna2023magnetic} including \frb. Thus, propagation effects have also been proposed, such as mode-mixing \citep{cheng1979theory} and generalised Faraday rotation (GFR) \citep{vedantham2019faraday}. Mode-mixing is discussed in Section~\ref{sec:magnetar} in more detail, and we focus first here on whether GFR can explain the observed properties of \frb.

In GFR, the propagation through a hot (relativistic) plasma (or admixture of strongly magnetised hot and cold, non-relativistic plasma) results in the natural modes of the plasma having either linear or elliptical polarisation. The modes have different refractive indices so travel at different speeds through the plasma. This can result in linear to circular polarisation conversion when recombined. We can model this effect using the phenomenological model described in \cite{lower2021dynamic}. In this model, the frequency-dependent Stokes $Q$, $U$ and $V$ parameters are projected onto the Poincare sphere as a polarisation vector $\rm \textbf{P}(\lambda)$, where $\rm \lambda$ is the wavelength of radio waves. GFR can be replicated on the Poincare sphere by introducing rotations on $\rm \mathbf{P}(\lambda)$ about the Stokes $U$ and $V$ axes, with angles of $\theta$ and $\phi$ respectively. This allows us to model the measured Stokes parameters without assuming any underlying physics. GFR will induce a frequency dependence on the linear polarisation of the form

\begin{equation}
\Psi(\lambda) = \Psi_{o} + \rm GRM (\lambda^\alpha - \lambda_{o}^\alpha),
\end{equation}

where $\rm \Psi(\lambda)$ is the PA due to GFR, GRM is the generalised Faraday rotation measure (an analogue of RM),  $\Psi_{o}$ is the intrinsic PA and $\alpha$ is the frequency exponent. $\alpha$ can then be used to infer the underlying physics of the circum-burst environment and/or nearby scattering regions. For example, dispersion in a highly relativistic plasma can induce strong GFR, with a frequency dependence as high as $\alpha = 3$ \citep{melrose1997response}. Detection of such effects would place significant new constraints on the FRB environment, and hence on the FRB progenitor.  

Conventional Faraday Rotation (FR) may also be induced by propagation through a cold magnetised plasma, and has a wavelength dependence $\alpha = 2$ (Section~\ref{sec:pol}). The conditions required for normal FR could exist anywhere along the FRB sight-line, but since any GFR would be expected to take place close to the source, we use the extended GFR-FR model described in \cite{Uttarkar:2024} to measure the shifts imposed on $\rm \mathbf{P}(\lambda)$ by both FR and GFR:

\begin{equation}
    \label{eq:FR-GFR}
    \rm \textbf{P}_{FR-GFR}(\lambda) = \textbf{R}_{\psi}\textbf{R}_{\theta \phi}\textbf{P}(\Psi, \chi),
\end{equation}

where $\rm \textbf{R}_{\psi}$ is the FR-induced rotation matrix, $\rm \textbf{R}_{\theta \phi}$ the GFR-induced rotation matrix and $\rm \textbf{P}_{FR-GFR}(\lambda)$ the measured polarisation vector. 

$\chi$ in Eq.~\ref{eq:FR-GFR} is the ellipticity angle that is used to model any intrinsic elliptical polarisation in the FRB:

\begin{equation}  
\chi = \frac{1}{2} \taninv\left(\frac{V(\nu)}{\sqrt{Q(\nu)^2+U(\nu)^2}}\right).
\label{eq:ellipse_ang}
\end{equation}

In order to constrain the presence of GFR in \frb, we model the frequency dependant polarisation of the two bright peaks in the first sub-burst B1 and B2 (as labelled in Fig.~\ref{fig:burstfit}) using this FR-GFR phenomenological model. For each peak we took a time-averaged window of data centred around the peak using the boxcar widths in Table~\ref{tab:GFR_table}. We applied a Gaussian filter as described in \citet{Price:2019} with a spectral window of 2 MHz and temporal window of 4 $\mu$s. The time-averaged Stokes $Q$, $U$ and $V$ spectra were used to constrain the parameters of the FR-GFR model. The posteriors for each peak are shown in Table~\ref{tab:GFR_table}. 

\begin{figure*}
        \subcaptionbox[width=1\columnwidth \label{subfig:pulse1}]{}
	{\includegraphics[width=1\columnwidth]{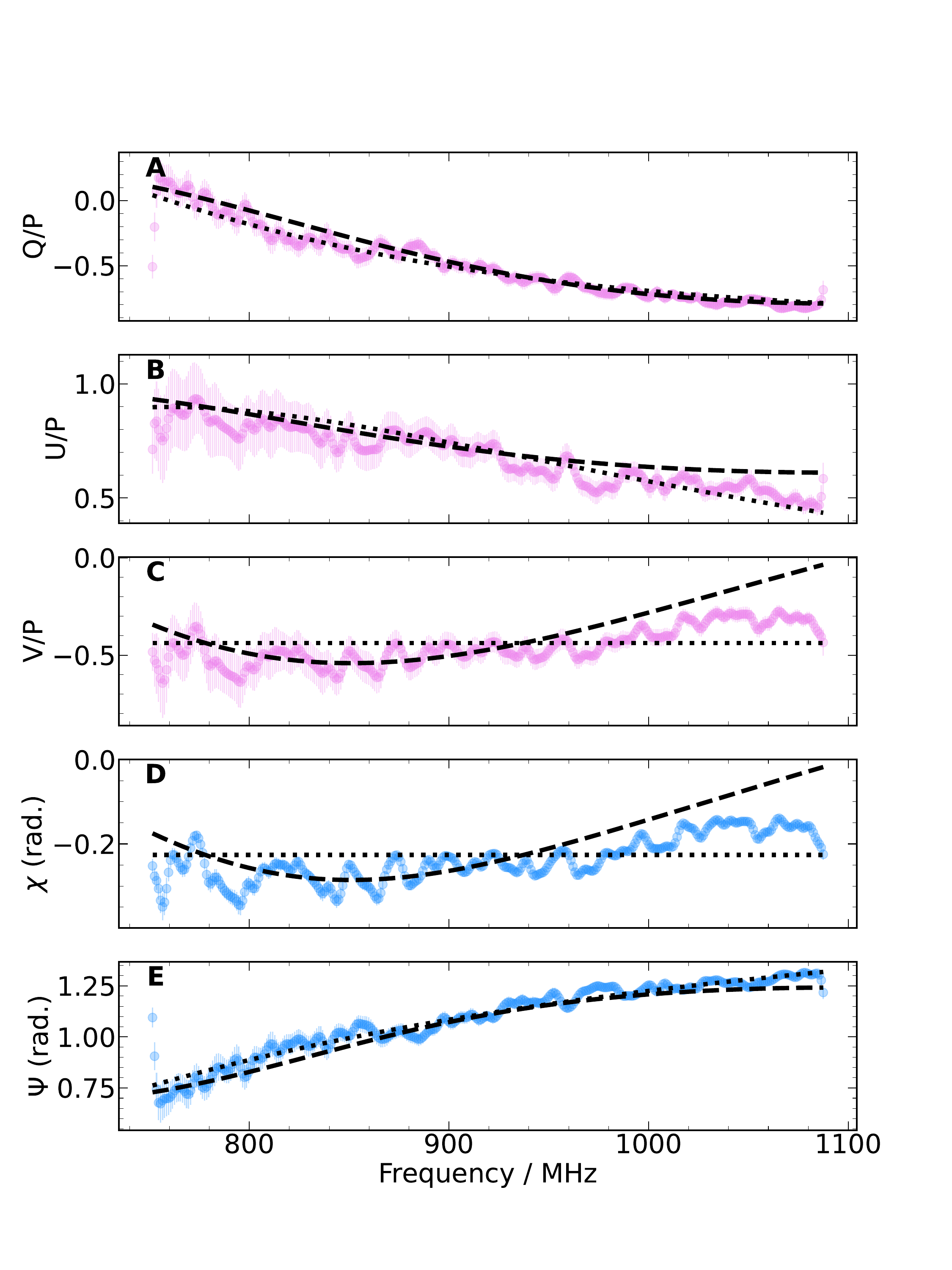}}
          \subcaptionbox[width=1\columnwidth \label{subfig:pulse2}]{}
	{\includegraphics[width=1\columnwidth]{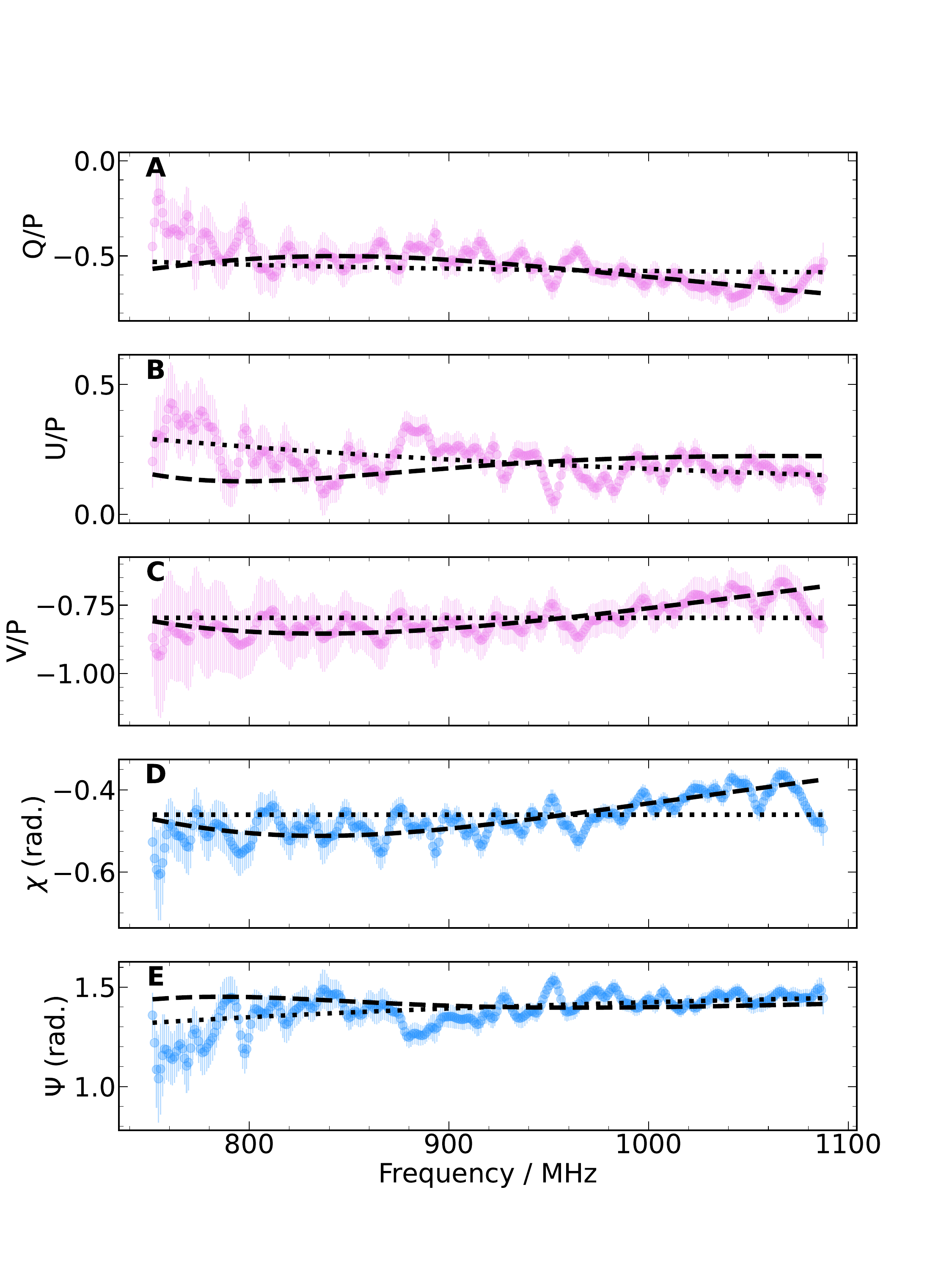}}
       
       \caption{The FR-GFR and FR fit to FRB 20230708 polarisation profile of the two bright peaks in the first sub-burst B1 (panel (a)) and B2 (panel (b)). The time averaged data for Stokes-Q, U, and V for FRB 20230708 are shown in pink points in Panels A, B, and C. The EA and PA variations across the frequency are shown in blue points in Panels D and E. The ellipticity angle ($\chi$) and the positional angle ($\Psi$) are shown in Panels D and E in blue points. We show the FR model fit for the data in black dotted lines. The black dashed lines show the FR-GFR model fit to the data. }
       \label{fig:QUVFIT}         
\end{figure*}

In Table~\ref{tab:GFR_table} for both peaks, the parameter $\alpha$ is consistent with the lower prior boundary at $\alpha = 0$, meaning we cannot detect $\alpha$, only an upper limit. Additionally, the FR-GFR modelling seen in Fig.~\ref{fig:QUVFIT} seems to recover some frequency-dependent change between linear to circular polarisation. However, neither the FR nor FR-GFR models provide a  good fit to the observed polarisation properites of the burst, implying the frequency-dependent changes in polarisation are either intrinsic to the emission mechanism, or a result of coherent/partially coherent mixing between orthogonally polarised modes \cite[e.g.][]{oswald2023pulsar2}. In any case, the upper limits on $\alpha$ in the FR-GFR model imply that generalised Faraday rotation in a relativistic plasma is not the primary origin of the observed CP.

\input{GFR_table}

\subsection{Total intensity spectra analysis}

 To create total intensity spectra of \frb, we reconstructed the Stokes $I$ dynamic spectrum with 100 kHz channel resolution and 10~$\mu$s time resolution. We used the first bright component of the burst due to the high S/N and applied a matched filter based on the model fit of the sub-burst described in Section~\ref{sec:time}.
 
 First, we searched for possible scintillation in \frb. By visual inspection, there does not seem to be any indication of scintillation as the spectrum looks smooth across the full band. Nonetheless, we used a method described in \citet{sammons2023two} and \citet{ocker2022large} to measure the scintillation bandwidth. To decouple scintillation from any frequency structure in the FRB, we took the brightest sub-burst and averaged in time to obtain frequency spectra. A fitted model of the spectrum was then subtracted to obtain a residual spectrum. A 3rd order polynomial was used as this was the simplest model that best replicated the broad frequency structure in the sub-burst. We then calculated the ACF, which we modelled to be a Lorentzian function 
\begin{equation}
    \mathrm{ACF}(\Delta \nu) = m^{2}\frac{\nu_{dc}^{2}}{\nu_{dc}^{2} + \Delta \nu^{2}},
    \label{eq:acf}
\end{equation}

\noindent
where $\Delta \nu$ is the frequency lag in the ACF and $\nu_{dc}$ and $m$ are the decorrelation bandwidth (i.e scintillation bandwidth) and modulation index \citep{macquart2019spectral}. 
The ACF was fitted using a simple least squares method.
We note that subtracting the broad frequency features of the sub-burst means that we are insensitive to broad scintillation bandwidths. However, we can not meaningfully differentiate between broad scintillation bandwidths and intrinsic frequency structure of the FRB. Measuring $\nu_{dc}$ over the full bandwidth shown in the ACF fit of Fig.~\ref{fig:scint} we derived values of $\nu_{dc}$ = 0.38 $\pm$ 0.07 MHz and $m$ = 0.31 $\pm$ 0.07 respectively. It bears noting that $\nu_{dc}$ is similar to the predicted NE2001 model for the scintillation bandwidth expected due to scattering in the local environment (i.e. the Milky Way) $\nu_{\rm NE2001}$ = 0.43 MHz. However, small-scale structure in the ISM that cannot be captured by Galactic electron density models such as NE2001 means that these predictions can frequently be in error by up to an order of magnitude \citep{sammons2023two}, so we treat this as a probable but not definitive detection of Galactic scintillation.

If the observed correlation at short frequency lags {\em is} due to scintillation, the decorrelation bandwidth should evolve with frequency across the observing band. We attempted to confirm this by performing a sub-band analysis using a similar method to that for the pulse broadening time described above. However, the signal-to-noise ratio was too low in the lower half of the band due to the (presumably) intrinsic fall off in the FRB brightness. Thus, limits on the frequency dependence of the decorrelation bandwidth were non-constraining.

Using the derived values of $\nu_{dc}$ and $\tau_{s}$ for the first sub-burst and assuming scintillation is present, we can describe the scintillation properties following the approach of \citet{sammons2023two}. First, we calculated the scattering constant $C$ using 

\begin{equation}
    C = 2\pi \nu_{dc}\tau_{s}
    \label{eq:scint_constant}
\end{equation}

\noindent
where $\nu_{dc}$ and $\tau_{s}$ are converted to units of Hz and s respectively, to obtain a value of $C$ = 395. Compared to similar CRAFT FRBs in \citet{sammons2023two}, the value of $C$ appears lower than the FRBs that were found to possess some measure of strong scintillation and higher than those that either did not or were inconclusive. However the value is much greater than unity, the value  expected if scattering and any putative scintillation originate in the same screen.

Assuming we removed all intrinsic frequency structure when deriving $\nu_{dc}$, it is feasible that \frb\ shows some level of scintillation. Again, it is difficult to confirm this because (1) we are unable to identify any visual cues of scintillation; and (2) we are also unable to measure a power law function describing this scintillation. Assuming \frb\ does show scintillation, we attempted to constrain a potential scattering environment near the FRB source. In \citet{sammons2023two} a dual thin-screen model was used to describe the effects of scattering, since a single thin-screen is insufficient to explain the scatter broadening and scintillation. The relationship between the relative distance to the thin-screen local to the FRB as measured from the source of the FRB $L_{x}$, and the thin-screen local to the Milky Way measured from us the observers $L_{g}$, can be described as

\begin{equation}
    L_{x}L_{g} \lesssim \frac{D_{s}^{2}}{2\pi \nu^{2}_{c}(1+z)} \frac{\nu_{dc}}{\tau_{s}}
    \label{eq:twoscreen},
\end{equation}

where $D_{s}$ is the luminosity distance between the FRB source and the observer at a redshift $z$ assuming a $\Lambda$-CDM cosmology with a Hubble constant and matter density of $H_{0}$ = 67.4 km/s/Mpc and $\Omega_{m}$ = 0.315 respectively \citep{aghanim2020planck}, and $\nu_{c}$ is the central frequency of the observing band. Using Eq.~\ref{eq:twoscreen} we estimated upper limits on $L_{x}L_{g} \lesssim$ 97 $\pm$ 17 kpc$^{2}$. We also estimated L$_{x}$ by assuming that the dominant galactic thin screen will be at a distance L$_{g} < z_{0}$, where $z_{0}$ = 1.57 $\pm$ 0.15 kpc is the scale height of the Milky Way derived from \citet{ocker2020electron}. Taking L$_{g}$ = 1.57 kpc we estimate an upper limit on the distance between the FRB progenitor and its nearest scattering screen to be L$_{x} \lesssim$ 62 $\pm$ 13 kpc. Thus, the extra-galactic thin-screen is likely to reside within the host galaxy halo, either within the CGM, ISM or circum-burst environment.

%% file: GFR_table.tex
\begin{table}

	\begin{center}
 
 \caption{The FR-GFR parameters for the two bright peaks in the first sub-burst of \frb. We show the 95\% confidence upper limit for the GFR and $\alpha$ parameters. }\label{tab:GFR_table}
		\begin{tabular}{c|c|c}
			\hline
			 & B1 & B2 \\
			\hline
                \thead{Boxcar width \\ ($\mu$s)} &  380     & 280 \\
			\thead{RM \\ (rad m$^{-2}$)} & -3.01$_{-0.77}^{+1.95}$ & -13.08$_{-4.82}^{+4.28}$ \\
            \thead{GRM \\ (rad m$^{-\alpha}$)} & $<$ 18.82 & $<$ 12.4 \\
			\thead{$\Psi$ \\ (deg)} & 75.4$_{-2.59}^{+3.05}$ & 77.27$_{-6.15}^{+4.95}$ \\
			\thead{$\chi$ \\ (deg)} & -32.92$_{-3.17}^{+10.68}$ & -18.0$_{-5.35}^{+24.84}$ \\
			\thead{$\phi$ \\ (deg)} & 35.02$_{-6.74}^{+11.78}$ & 136.11$_{-22.98}^{+23.5}$ \\
            \thead{$\alpha$ \\ } & $<$ 2.08 & $<$ 1.16 \\
			\thead{$\theta$ \\ (deg)} & 81.55$_{-6.3}^{+21.36}$ & 22.66$_{-10.71}^{+51.49}$ \\
            \hline
		\end{tabular}

	\end{center}

\end{table}



%% file: Discussion.tex
\section{Discussion}
\label{sec:DISCUSSION}

The temporal structure of \frb\ is suggestive rather than definitive evidence for the presence of periodicity. The 1.77$\sigma$ evidence is comparable to most other FRBs for which quasi-periodicity has been mooted (such as FRBs 20210206A and 20210213A \citep{chime2022sub} and FRB 20201020A \citep{Pastor_Marazuela_2023}); only FRB 20191221A \citep{chime2022sub} shows clearer 6.5$\sigma$ evidence for periodicity. While more robust methods of generating a null hypothesis test that accounts for component sub-structure could improve the measured evidence for \frb, it is doubtful it would be definitive of periodicity. 

Interestingly, \frb\ shows an extremely shallow scattering index of $\alpha = -2.17$. Pulsar emission at low frequencies has shown similar values \citep{geyer2017scattering} using the isotropic thin-screen model we have adopted in this analysis. Models using anisotropic scattering and/or finite scattering screens have shown results slightly more consistent with theoretical predictions of Kolmogorov scattering \citep{geyer2017scattering}. However, we do not attempt to resolve this here as these methods are unlikely to be constraining given the limited bandwidth and lack of repeat bursts from \frb\ that could help infer the physics of the scattering region. Additionally, the location of the turbulent plasma responsible for the temporal broadening of the FRB emission is constrained to lie within $\lesssim$ 62 kpc of the source of the burst, within the host galaxy halo. The lack of a lower limit on the screen distance means that the observed temporal broadening does not well constrain the scattering physics and the FRB progenitor. 

The strongest constraints on the progenitor of \frb\ come from the observed polarisation, as the burst components show a variety of polarisation properties. The majority of the known quasi-periodic FRBs reported in the literature do not have full polarisation data with one exception. FRB 20210206A \citep{chime2022sub} shows extremely high fractional linear polarisation with a relatively flat PA profile, and little CP. \frb\ on the other hand shows a plethora of polarisation features. The FRB shows a structured PA profile across its burst, starting with a shallow PA sweep of $\simeq$ 30$^{\circ}$ across the first bright sub-burst shown in Fig.~\ref{fig:comp1_fit}, followed by a shallow decline across the rest of the burst. The bright sub-burst also exhibits apparent high time-dependant CP conversion peaking at $V$/$I\simeq 75\%$ in the trailing peak (see Fig.~\ref{fig:comp1_fit}), along with a change of handedness. While this temporal structure is suggestive of propagation effects such as GFR, frequency-dependent modelling was not consistent with what is predicted for propagation through a relativistic plasma. The limited burst bandwidth makes it difficult to explore other propagation effects that could explain all the polarisation features we see in the first bright component.

Given the constraints on these properties, we will now examine popular FRB progenitor models for \frb.

\subsection{Rotation powered pulsar}

The first progenitor to examine for \frb\ is the rotating pulsar. Given the apparent quasi-periodicity of $T = 7.267$ ms across the FRB, this would sit firmly in the milli-second pulsar (MSP) population. However, the polarisation properties of the first sub-burst component differs substantially from the rest of the burst. The first bright component shows a small $\simeq$ 30$^{\circ}$ PA sweep and significant apparent CP conversion (see Fig.~\ref{fig:comp1_fit}). There is also a slight CP handedness change from the first bright peak to the second. The remaining sub-burst components show flat PA profiles and consistently high LP. Pulsars have shown a variety of polarisation features similar to \frb\, showing flat/steep PA swings, CP conversion and handedness change \citep{oswald2023pulsar1}. 

While pulsars have even shown substantial variations in successive single pulses \citep{johnston2024thousand}, the specific configuration that would be implied if \frb\ were a pulsar with a millisecond spin period, namely, an interconnected PA wander from pulse to pulse, has not been observed.
\frb\ also shows broad emission in what would be pulse phase, whereas, in general, the rapid variations in polarisation properties seen in pulsars has been confined to narrow regions in their pulse profile. Overall, the polarisation seen in \frb\ disfavours a $\sim$ 7ms period pulsar.

\subsection{Magnetar}
\label{sec:magnetar}
An alternative explanation for the apparent quasi-periodicity in FRBs has been put forward by \citet{kramer2023quasi}, whereby the burst is related to microstructure in beamed emission of a slow rotating NS. The galactic population of NS (including pulsars and magnetars) have shown microstructures in their bursts \citep{kramer2007polarized, dai2019wideband}. In fact, both populations of pulsars and radio-loud magnetars have been shown to follow a unique relationship between the quasi-periodicity in their emission sub-structure, and the global rotational period of the NS \citep{kramer2023quasi}:

\begin{equation}
    \tau_{\mu} \simeq 10^{-3}T_{\rm NS}
\end{equation}

\noindent
where $\tau_{\mu}$ is the sub-structure periodicity and $T_{\rm NS}$ is the rotational period of the NS. Given the many apparent similarities between FRBs and NS, \citet{kramer2023quasi} suggests FRBs may follow a similar trend (although there has been no way to confirm the spin frequency of any putative FRB progenitors to date). Using this relationship and assuming $\tau_{\mu}$ is 7.267 ms, we would expect a hypothetical NS spin period of O(10) s for \frb, which would be consistent with the spin periods of known radio-loud magnetars. 

The PA profile is consistent with expectations based on observations of magnetar single pulses. There are very few studies of magnetar single pulses at such high time resolution, but it has been found that the PA of the individual peaks in the sub-structure of these single pulses often closely track the overall PA sweep seen in the integrated pulse profile \citep{kramer2007polarized, levin2012radio}. That is to say, the sub-structure shows an ordered, interconnected PA profile, which is what we see in \frb.

The first bright sub-burst of \frb\ shows a shallow $\simeq$ 30$^{\circ}$ PA sweep along with high CP and significant time-dependant CP conversion, which as mentioned above, could be indicative of propagation effects within a highly magneto-ionised environment such as mode-mixing \citep{cheng1979theory} or GFR \citep{lower2024linear}. GFR has been observed in radio emission along with significant time-dependant CP conversion, such as emission from the magnetar XTE J1810$-$197 \citep{lower2024linear} and even in one of the bursts from the repeating FRB 20201124A \citep{kumar2023propagation, kumar2022circularly}. In the case of \frb, however, the frequency dependence of polarisation does not support the presence of GFR. Mode-mixing in pulsars have shown a variety of polarisation properties, including PA jumps, high CP, handedness change in CP and depolarisation at higher frequencies \citep{oswald2023pulsar1, oswald2023pulsar2}. Similar mode mixing with orthogonal polarisation modes (OPM) has been seen in magnetar emission \citep{lower2021dynamic}. Considering the partial mode mixing model discussed in \citet{oswald2023pulsar2} some of the polarisation properties in \frb\ could be replicated, most notably the high CP if the coherent mixing fraction and phase delay between the two modes were sufficiently high, in addition to other properties such as the handedness change. However, it is difficult to test these models given the limited bandwidth and single burst for \frb.
Overall, many of the properties of \frb\ seems to better support a magnetar progenitor model as opposed to a spin down powered NS such as a millisecond pulsar.

It is also possible that the source of \frb\ is a standard long period pulsar . The sub-structure in pulsar emission has also been shown to possess apparent flat (or shallow) PA sweeps \citep{mitra2015polarized}. However, the NS spin down luminosity is orders of magnitude too low to power the FRB radio emission. 

\subsection{Coalescence of compact binary objects}
There are a number of proposed progenitor and emission mechanism models in the literature including maser models \citep{lyubarsky2014model,metzger2019fast}, cosmic combs \citep{yang2018bunching}, stable binary systems \citep{mottez2020repeating} in addition to those previously mentioned. Whether a subset of FRBs truly only occur once is still a topic of debate. If a class of genuine non-repeaters does exist, they could originate from cataclysmic events. One such scenario that could explain quasi-periodicity in FRBs is the merging of two compact objects in a binary orbit. In this scenario, one of the compact objects is highly magnetised (likely a NS) whilst the other (whether a BH or NS) acts as a perfect conductor. As the two objects coalesce, the companion of the magnetised object orbits through its strong magnetic field and acts  as a conductor which drives a current loop between the two objects in a unipolar inductor process \citep[e.g.][]{hansen2001radio,mcwilliams2011electromagnetic,piro2012magnetic,lai2012dc,wang2016fast, wang2018pre}. The motion of this companion object will result in an induced electric field and charged particles will be accelerated along the field lines which may result in FRB emission. As the orbital separation decreases, the separation between radio bursts decreases also in a predictable manner. In \cite{cooper2023pulsar} it is also expected that the energy reservoir available to produce radio emission is also strictly increasing. Thus, the brightness of successive bursts would also increase. Immediately we can see that this model is disfavoured due to the decrease in brightness of successive components across \frb.
 Any potential progenitor theory must also account for the unique polarisation properties observed with \frb, namely the brief high CP and PA profile. If we assume that near complete linearly polarised coherent curvature radiation is produced through the breaking of the current loop between the two orbiting bodies in this model, it is unclear where the circular polarisation would originate from. Perhaps the breaking of the current loop occurs within the magnetosphere of the highly magnetised NS, in which case the aforementioned propagation effects could be proposed. To our knowledge, this model has not been extended to explore the polarisation of the resulting FRB emission to date.

%% file: Conclusion.tex
\section{Conclusions}
\label{sec:CONCLUSION}

In this paper, we presented \frb, an FRB discovered by ASKAP through the CRAFT survey. We have analysed its rich morphology in detail, showing a number of components separated by a potential periodicity of T = 7.267 ms with a significance of 1.77$\sigma$. The mostly interconnected PA profile seen across the multiple components of this burst supports the emission representing micro structure from the beamed emission of a magnetar. The first bright sub-burst shows a number of polarisation features including high CP, temporal CP conversion, handedness change and a shallow PA sweep which could be the result of propagation effects, however, it is unlikely due to GFR.

Repetition searches of this FRB source could help further support the magnetar model given the repeating nature of the known radio-loud magnetar population, as well as help further constrain the temporal and polarimetric properties we observe for \frb. In addition, a number of FRB progenitor formation models predict persistent radio sources (PRS). Given our constraints on the temporal properties of \frb\ and the relatively nearby host galaxy, deeper radio continuum imaging may aid in identifying a potential PRS and further constrain the progenitor.

%% file: other.tex
\section*{Acknowledgements}

This scientific work uses data obtained from Inyarrimanha Ilgari Bundara / the Murchison Radio-astronomy Observatory. We acknowledge the Wajarri Yamaji People as the Traditional Owners and native title holders of the Observatory site. CSIRO’s ASKAP radio telescope is part of the Australia Telescope National Facility (https://ror.org/05qajvd42). Operation of ASKAP is funded by the Australian Government with support from the National Collaborative Research Infrastructure Strategy. ASKAP uses the resources of the Pawsey Supercomputing Research Centre. Establishment of ASKAP, Inyarrimanha Ilgari Bundara, the CSIRO Murchison Radio-astronomy Observatory and the Pawsey Supercomputing Research Centre are initiatives of the Australian Government, with support from the Government of Western Australia and the Science and Industry Endowment Fund.
Based on observations collected at the European Southern Observatory under ESO programme
108.21ZF. 
RMS and PAU acknowledge support through ARC Future Fellowship FT19010155. RMS and ATD acknowledge support through ARC Discovery Project DP220102305.
 MG is supported by the Australian Government through the Australian Research Council Discovery Project DP210102103.
MEL acknowledge support from the Royal Society International Exchange grant IES\textbackslash R1\textbackslash 231332.

\section*{Data Availability}
The data used in this work will be made available upon reasonable request.

\section*{Code}
The majority of the data analysis and plotting in this work used the ILEX python package for FRB analysis, which can be accessed at https://github.com/tdial2000/ILEX.

%% file: appendix.tex
\section{Additional plots}

\begin{figure*}
    \centering
    \includegraphics[width = 0.9\textwidth]{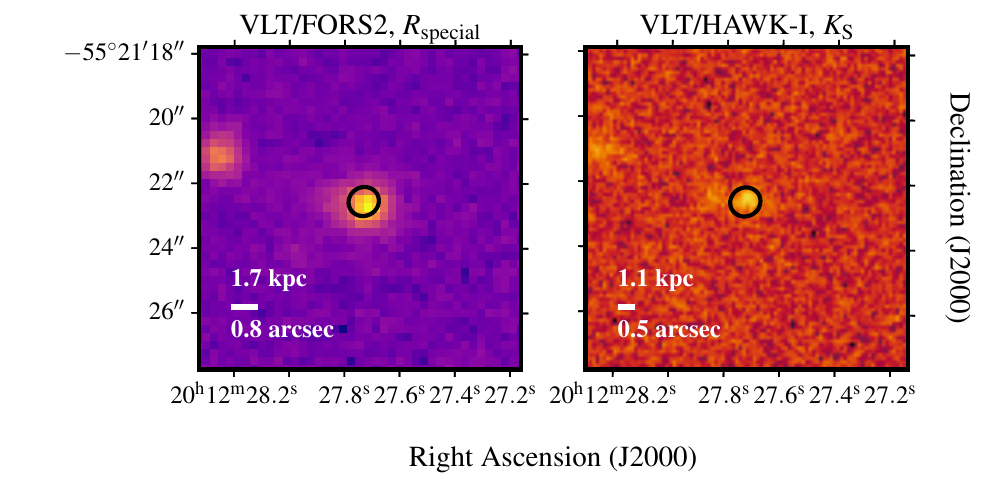}
    \caption{Left Panel: VLT FORS2 $R$-band image of the field around \frb\ with seeing of $0\farcs7$ used for the PATH analysis. Right Panel: VLT HAWK-I adaptive optics (AO) $K_s$-band imaging with delivered image quality of $0\farcs5$. The 1-$\sigma$ localisation of \frb\ is given by the black ellipse.}
    \label{fig:hawk}
\end{figure*}

\begin{figure}
    \centering
    \includegraphics[width = \linewidth]{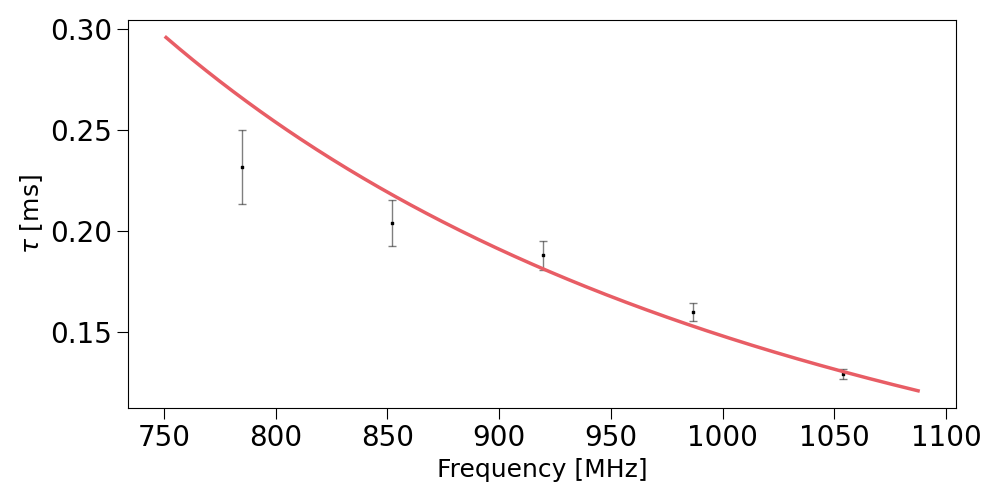}
    \caption{Scattering timescale vs frequency for first bright sub-burst. The solid red curve shows best fit power-law function $\tau = Af^{\alpha_{t}}$ for $\alpha_{t}$ = -2.17 where $A$ is a constant sampled along with $\alpha_{t}$.}
    \label{fig:specindex}
\end{figure}

\begin{figure}
    \centering
    \includegraphics[width = \linewidth]{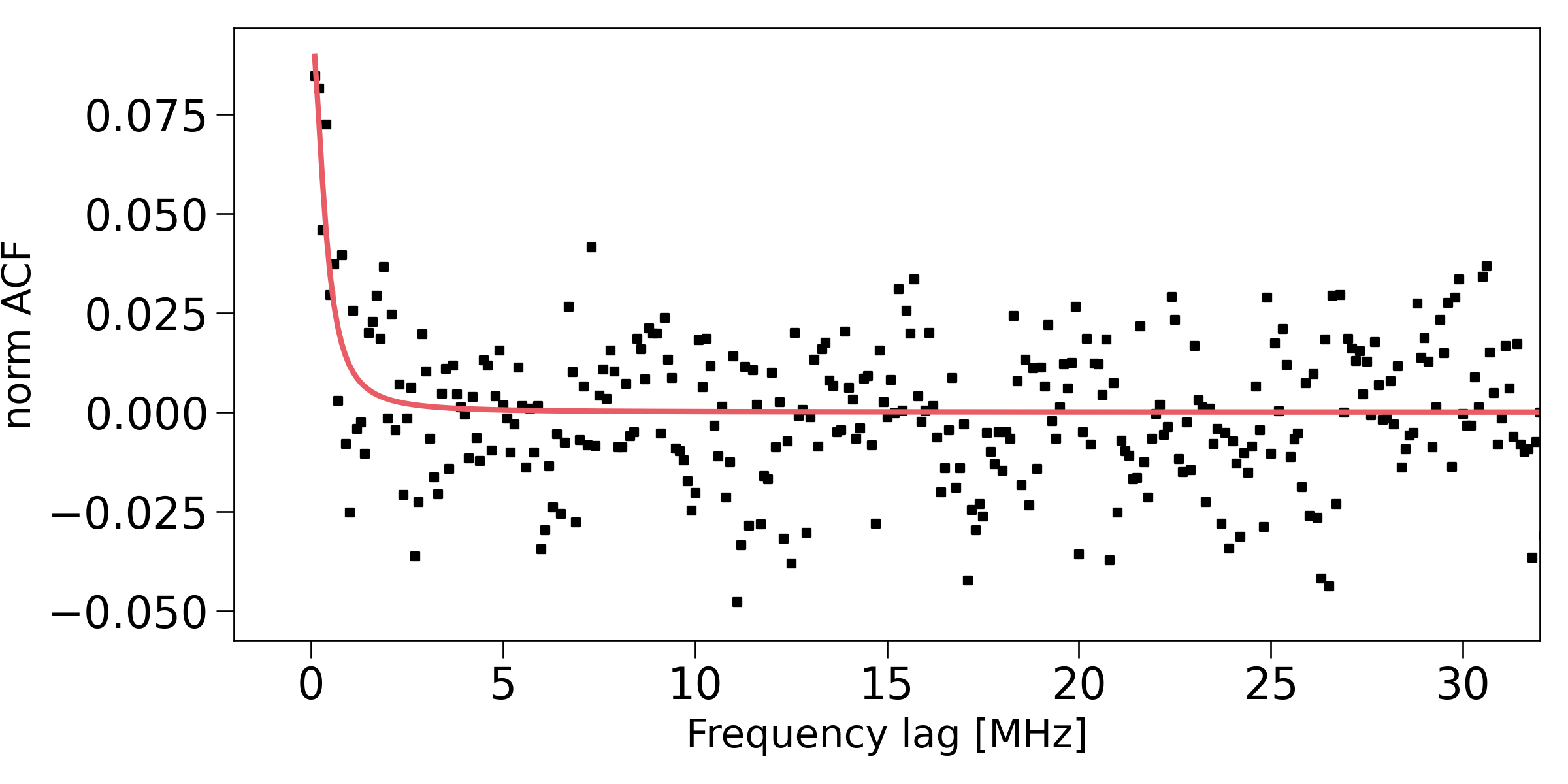}
    \caption{Normalised ACF residuals of main sub-burst spectra fitted against a Lorenztian function shown as a solid red line. The X-axis describes the frequency lag in MHz.}
    \label{fig:scint}
\end{figure}

\begin{figure}
    \centering
    \includegraphics[width = \linewidth]{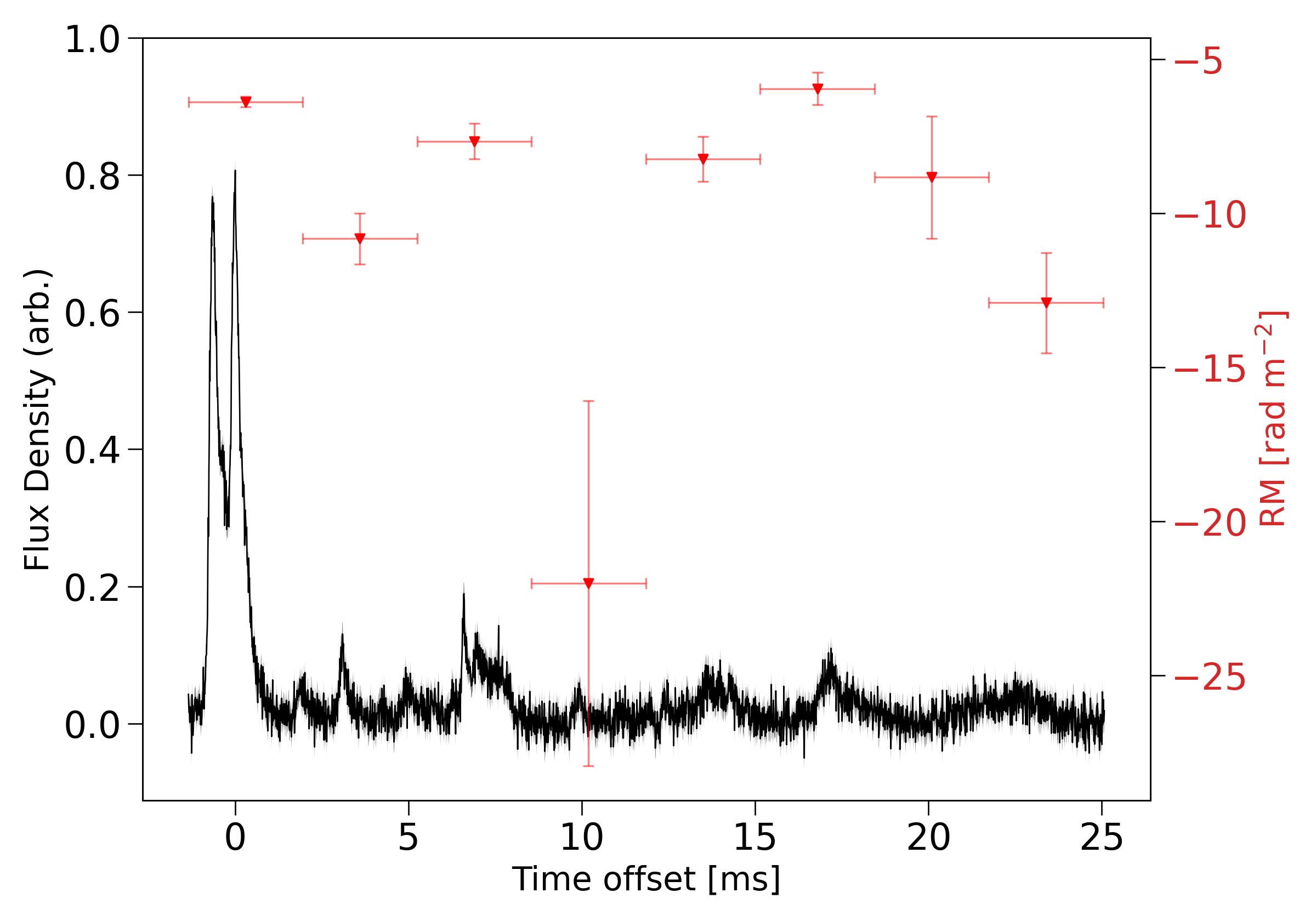}
    \caption{Measured RM (shown in red) across stokes $I$ time series burst of \frb\ (solid black line). Horizontal error bars show the region of data averaged in time to get spectra for RM fitting.}
    \label{fig:RM_burst}
\end{figure}

\begin{figure}
    \centering
    \includegraphics[width = \linewidth]{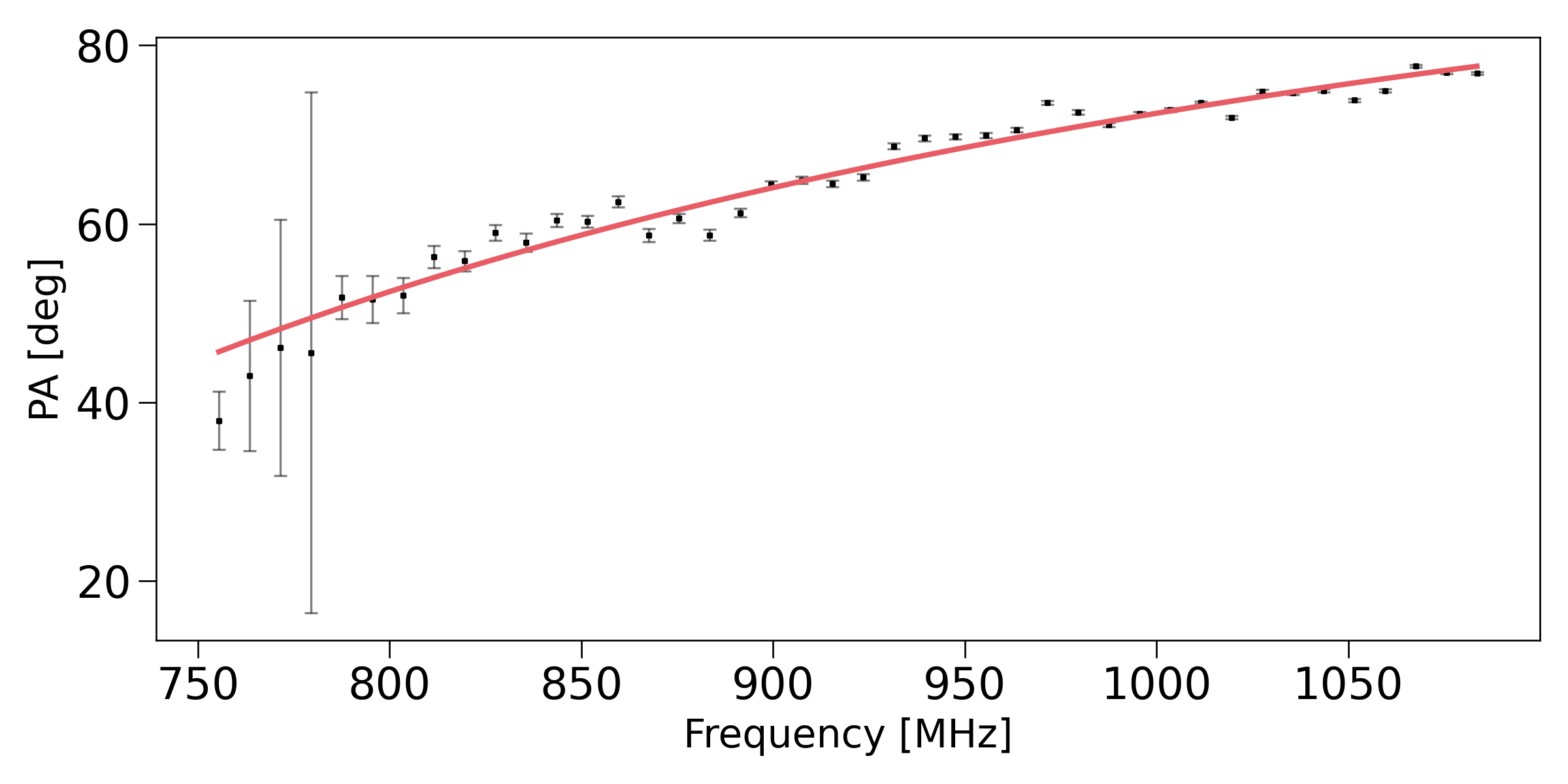}
    \caption{Plot of PA against frequency. The Measured PA shown in the black points was calculated using Eq.~\ref{eq:PA}, whilst the expected model PA shown in the solid red curve was calculated using Eq.~\ref{eq:RM} with RM = --6.90~rad/m$^{2}$.}
    \label{fig:RM_full}
\end{figure}

\section{Polarisation fractions}

The total polarisation $P(t)$ is

\begin{equation}
    P(t) = \sqrt{L_{\rm debias}^{2}(t) + V^{2}(t)}.
    \label{eq:total_pol}
\end{equation}

The continuum added polarisation fractions $\overline{l}$, $\overline{|v|}$ and $\overline{p}$ are calculated using \citep{oswald2023pulsar1}

\begin{equation}
    \begin{split}
    & \overline{l} = \frac{\sum{L_{\rm debias}(t)}}{\sum{I(t)}} \\
    & \overline{|v|} = \frac{\sum{|V(t)|_{\rm debias}}}{\sum{I(t)}} \\
    & \overline{p} = \frac{\sum{P(t)}_{\rm debias}}{\sum{I(t)}},
    \end{split}
    \label{eq:int_polfrac}
\end{equation}

where $|V(t)|_{\rm debias}$ needs to be debiased because we are taking the modulus of $V(t)$ \citep{karastergiou2003v, oswald2023pulsar1}. $P(t)_{\rm debias}$ is debiased in a manner similar to $L(t)_{\rm debias}$ using Eq.~\ref{eq:L_debias}.